\documentstyle[12pt]{article}
\begin{document}
\begin{center}
{\Large \bf A new conjecture extends\\ the $GM$ law
for percolation thresholds \\ to dynamical situations}
\vskip 0.6cm
{\bf Serge Galam\footnote{e-mail: galam@ccr.jussieu.fr}} \\
Laboratoire des Milieux D\'esordonn\'es et H\'et\'erog\`enes, Tour 13,
Case 86, \\ 4 place Jussieu, 75252 Paris Cedex 05, France

\vskip 0.6cm
{\bf Nicolas Vandewalle\footnote{e-mail:vandewal@gw.unipc.ulg.ac.be}}
\\
SUPRAS, Institut de Physique B5, Universit\'e de Li\`ege, \\
B-4000 Li\`ege, Belgium

\vskip 0.6cm
Int. J. Mod. Phys. C \underline{9} (1998) 667-671

\end{center}

\vskip 3.0cm

\begin{abstract}
The universal law for percolation thresholds proposed by Galam and Mauger
($GM$) is found to apply also to dynamical situations. This law depends
solely on two variables, the space dimension $d$ and a coordinance number
$q$. For regular lattices, $q$ reduces to the usual coordination number
while for anisotropic lattices it is an effective coordination number. For
dynamical percolation we conjecture that the law is still valid if we use
the number $q_2$ of second nearest neighbors instead of $q$. This
conjecture is checked for the dynamic epidemic model which considers the
percolation phenomenon in a mobile disordered system. The agreement is good.
\end{abstract}

\vskip 2.0cm
keywords: percolation threshold --- epidemic model --- lattice --- tree
\newpage
{\noindent \large 1. Introduction}
\vskip 0.6cm

Recently, a universal power law for both site and bond percolation
thresholds was postulated by Galam and Mauger ($GM$) \cite{galam1,galam2}.
The $GM$ formula is
\begin{equation}
p_c^{GM} = p_0 { \left[ (d-1) (q-1) \right] }^{-a} d^b
\end{equation} with $d$ being the space dimension and $q$ a coordinance
variable. For regular lattices, $q$ is the lattice coordination
number. The exponent $b$ is either equal to $a$ for bond percolation
or to $0$ for site percolation. Three classes characterized by three
different sets
of parameters $\{ a; p_0 \}$ were found. The first class includes
two-dimensional triangle, square and honeycomb lattices with $\{
a=0.3601; p_0=0.8889 \}$ for site percolation and $\{ a=0.6897; p_0=0.6558
\}$ for bond percolation. Two-dimensional Kagom\'e and all (hyper-)cubic
lattices in $3 \le d \le 6$ constitute the second class with $\{
a=0.6160;p_0=1.2868 \}$ and  $\{ a=0.9346; p_0=0.7541 \}$ for site and bond
respectively. The third class corresponds to high dimensions ($d > 6$).
There, $b=2a-1$ and $p_0=2^{a-1}$ with $a=0.8800$ for site and $a=0.3685$
for bond percolation. For hypercubic lattices ($q=2d$), the
asymptotic limit leading term is identical to the Cayley tree threshold.
For both site and bond
percolation \cite{percol} it is,
\begin{equation}
p_c = \frac{1}{z}
\end{equation} where $z$ is the branching rate of the tree, i.e. $z\equiv q-1$.

In high dimensions, it has been reported that more $GM$ classes should be
taken into account \cite{vanm}. Extension to anisotropic and aperiodic
lattices has also been also reported \cite{galam2}. The formula (1) remains
valid. However to preserve
its high accuracy $q$ should be replaced by an
effective non-integer value $q_{eff}$ which is different from a simple
arithmetic average.

In this letter, making a simple conjecture we extend the
validity of the $GM$ formula (Eq.(1)) to dynamical situations like epidemics or
contagion models. The results obtained fit well
to available data.

\vskip 1.0cm
{\noindent \large 2. Epidemic models}
\vskip 0.6cm

The epidemic model \cite{epidemic} considers the Eden growth \cite{eden} of
a phase in a disordered medium with static impurities. In this model, a
fraction $x$ of
particles are randomly dispersed on a lattice. These particles act as
hindrances for the random growth of a spreading phase. For $x < x_c$, the
cluster is growing forever while for $x > x_c$ the growth of the cluster is
stopped after a finite number of timesteps. This unblocked-blocked growth
transition is closely related to a site percolation phenomenon since $x_c =
1 - p_c$.

Recently, the dynamical epidemic model \cite{jphysa,dynepid} has been
introduced
in order to provide a phenomenological basis for the aggregation of mesoscopic
impurities along crystal growth surfaces and interfaces \cite{nickpre}. In
the dynamic
epidemic model, the particles are mobile hindrances for the random growth
of a phase. When the growth front reaches a particle, the latter is supposed to
move to an empty nearest neighboring site in order to minimize its contact
with the spreading phase if such a move is possible. This rule is
illustrated in Figure 1. This
repulsive dynamical interaction leads to an aggregation phenomenon along
the growth interface and leads further to some reorganization of the mobile
medium \cite{dynepid}. The threshold $p_c$ for dynamical situations is
quite different from the static medium. The formation of growth
instabilities along the
propagating front has been recently reported \cite{nickpre} to be the
source of the organization process.

The dynamic epidemic model has been numerically studied at $d=2$
\cite{jphysa,dynepid} and $d=3$ \cite{bulgares}. Moreover, an exact
solution
has been obtained on tree-like structures which are illustrated in Figure
2. The percolation thresholds for either static or dynamic epidemics are
listed in Table I as well as the thresholds predicted by the $GM$ formula.

\vskip 1.0cm
{\noindent \large 3. The conjecture}
\vskip0.6cm

Let us consider first a square lattice. As described above and
illustrated in Figure 1, when a particle is reached by the interface, the
particle tries to move to another neighboring empty site as in avalanche
processes in sandpiles. If the
neighborhood of the particle is completely occupied, the particle remains
static and the cluster growth is pinned there. Intuitively, one can
consider that the unblocked-blocked growth transition depends at least on
the occupation of
sites in the front neighborhood including the $q_2$ second neighboring
sites (neighbors of neighbors). Not to be mistaken with next nearest
neigbhors.
The extended neighborhood of a site is shown in Figure 3. It is worth
noticing that on regular lattices
\begin{equation}
q_2 = d q.
\end{equation} Taking $q_{eff}=q+q_2=12$, as for percolation with next nearest
interactions \cite{galam1}, the $GM$ law (Eq. (1), site first class)
does not hold and does not reproduce the
dynamical threshold, for instance $p_c=0.44$ for squares. Nevertheless,
one can notice that taking
$q_{eff}=q_2=8$
provides the value $0.4411$ for $p_c$ which is ``exact" within the two
digits of available data (see Table I).

Using the same substitution, the universal law \cite{galam1}
holds true also for the cubic ($d=3$) case
for which
$q_2=18$ and $p_c=015$. It yields (for site second class)
$p_c=0.1466$ which is again ``exact". These results ground the
following conjecture: Percolation properties of a dynamic epidemic
process are identical to the associated static ones with
substituting $q_2$ to $q$.

We are now in a position to check the above conjecture in the case
of the Cayley tree for which the thresholds are known exactly.
The relevant parameter is the branching rate $z$ instead of $q$.
The parameter $z$ can be considered as the number of nearest neighboring
sites $\tilde q$ towards the tree extremities. In fact, there is no
distinction between directed percolation and simple percolation on a
tree-like structure.
The exact solution of the dynamic epidemic model on a Cayley tree
\cite{cayley} is
\begin{equation}
p_c=1/z^2.
\end{equation} The number $\tilde q_2$ of second
neighboring sites towards the leaves is $z^2$. Thus from Eq.(2), this
theoretical result corroborates the above conjecture that $q$ should be
replaced by $q_2$ in the $GM$ universal law.

On the trees decorated with loops like those illustrated in Figure
2b and 2c the sites are non-equivalent. An effective number $\tilde q_2$ of
second sites towards the leaves
can be calculated by averaging $\tilde q_2$ on the various sites of the
tree structure. Effective values of the number $\tilde q_2$ for trees
decorated with loops are given in Table I. Again, considering $\tilde q_2$
instead of $q$ in Eq.(2) provides a good value for $p_c$ in the case
of trees decorated with loops (see Table I). However, the agreement is not
as good as for the square and cubic lattices. This discreapancy is to be put
in parallel to the finding \cite{galam2} that for anisotropic lattices an
effective number of neigbhors should be used instead of the arithmetic average.

\vskip 1.0cm
{\noindent \large 4. Conclusion}
\vskip0.6cm

In this letter we have proposed a mapping of the percolation threshold for
a dynamical situation to a static one by making a simple conjecture
with respect to the relevant neigbhoring sites. The
conjecture states that the dynamical percolation case reduces to the static
percolation one considering that the percolation threshold should include
the effect of the second nearest neigbhors only.
This conjecture was then checked using the $GM$ universal law for percolation
thresholds. The results are very convincing. It was also found to be satisfied
in the Cayley tree case. More results on a larger lattice spectrum would allow
a more definite check of our conjecture.

\vskip 1.0cm

{\noindent \large Acknowledgements}
\vskip 0.6cm
NV thanks the FNRS for financial support. Thanks to M.Ausloos for fruitful
discussions.

\newpage
\vskip 2.0cm
\begin{center}
\begin{tabular}{|c|c|c|c|c|c|c|c|} \hline

Lattice & Ref. & $p_c$ static & $p_c^{GM}$ static & $p_c$ dynamic
& $p_c^{GM}$ dynamic & $q$ & $q_2$ \\ \hline
square ($d=2$) & \cite{jphysa,dynepid} & 0.5928 & 0.5985 & 0.44
& 0.4411 & 4 & 8 \\ \hline
cubic ($d=3$) & \cite{bulgares} & 0.3116 & 0.3115 & 0.15 & 0.1466
& 6 & 18 \\ \hline  \hline
Tree & Ref. & $p_c$ static & $1/\tilde q$ & $p_c$ dynamic
& $1/ \tilde q_2$ & $\tilde q$ & $\tilde q_2$ \\ \hline
Cayley tree & \cite{cayley} & $1/z$ & $1/z$ & $1/z^2$ & $1/z^2$
& $z$ & $z^2$ \\ \hline
tree ($\triangle$ loops) & \cite{treepre} & 1/2 & 1/2 & 0.151 & 0.1667
& 2 & 18/3 \\ \hline
tree ($\Box$ loops) & \cite{treepre} & 0.597 & 0.625 & 0.269 & 0.2778
& 8/5 & 18/5 \\ \hline
\end{tabular}
\end{center}
\vskip 1.0cm
{\noindent \bf Table I} --- The thresholds $p_c$ for both static and
dynamic epidemic models compared to the associated $p_c^{GM}$ thresholds
from Eq.(1). All lattices studied up to now to our knowledge are listed
with pertinent
references. The number of nearest neighboring sites $q$ as well
as the second neighborhood $q_2$ are also given. Trees are as those shown in
Figure 2.

\newpage
{\noindent \large Figure captions}

\vskip 1.0cm
{\noindent \bf Figure 1} --- Illustration of one growth step of the dynamic
epidemic model. The growing phase is drawn in white, the mobile particles
(hindrances) are drawn in black and the empty sites are shown in grey: (a)
one empty site denoted by a cross in contact with the spreading cluster is
selected at random; (b) the growth takes place there and the particle
touched by the new unit jumps towards a neighboring site in order to reduce
its contact with the cluster.

\vskip 1.0cm
{\noindent \bf Figure 2} --- Three different trees with branching rate $z=2$
as discussed in this paper: (a) Cayley tree, (b) tree decorated with
triangular loops, and (c) tree decorated with square loops.

\vskip 1.0cm
{\noindent \bf Figure 3} --- Nearest neighbors and second neighbors on a
square lattice. The $q_2=8$ second neighbors which play the fundamental
role in the dynamic epidemic model are denoted in grey.

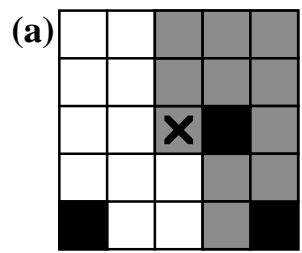
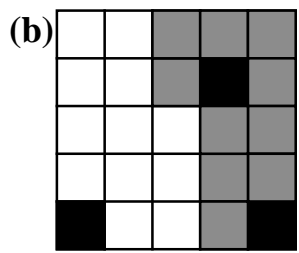
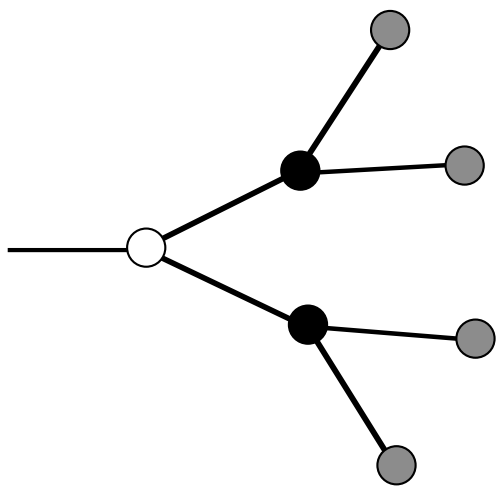
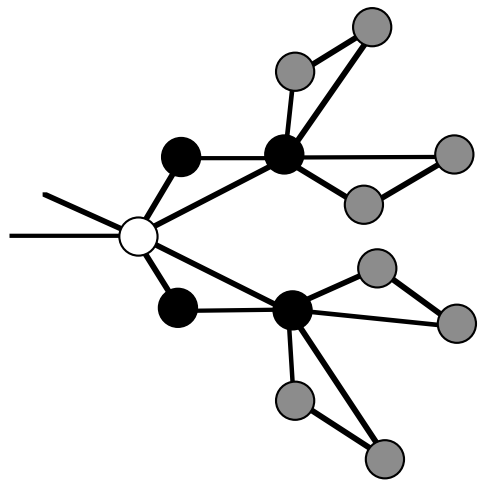
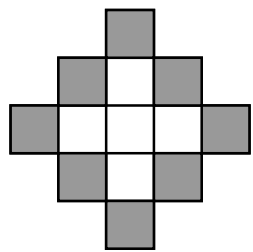
\end{document}